\documentclass[a4paper,12pt, epsfig]{article}
\usepackage{epsfig}

\usepackage{graphicx}
\usepackage{caption}
\usepackage{float}
\usepackage{subfigure}
\usepackage{braket}
\usepackage{amsmath}
\usepackage{url}
\usepackage[normalem]{ulem}
\usepackage[style=numeric,sorting=none]{biblatex}
\def\NvDEx{N$\nu$DEx}
\def\82SeF6{$^{82}$SeF$_6$}
\def\Se82{$^{82}$Se}
\def\SeF6{SeF$_6$}
\def\0vbb{0$\nu\beta\beta$}
\def\2vbb{2$\nu\beta\beta$}

\begin{document}
\def\thefootnote{\fnsymbol{footnote}}
\def\thetitle{Neutron Activation Background in the \NvDEx~Experiment}
\def\autone{Qianming Wang}
\def\auttwo{Zeyu Huang}
\def\autthree{Pengchong Hu}
\def\autfour{Emilio Ciuffoli}
\def\affa{Lanzhou University, Lanzhou 730000, China}
\def\affb{Institute of Modern Physics, NanChangLu 509, Lanzhou 730000, China}

\begin{center}
{\large {\bf \thetitle}}

\bigskip

\bigskip

{\large \noindent  \autone{${}^{1}$}, \auttwo{${}^{1}$}, \autthree{${}^{1}$} and \autfour{${}^{2}$}\footnote{emilio@impcas.ac.cn}
}

\vskip.7cm

1) \affa\\
2) \affb\\

\end{center}

\begin{abstract}

      { 
        An extremely low-background environment is a crucial requirement for any neutrinoless double beta decay experiment. Neutrons are very difficult to stop, because they can pass through the shields and activate nuclei in the detector, even inside the fiducial volume itself. Using Geant4 simulations we have studied the neutron background for \NvDEx-100 and the most efficient way to reduce it. Using a 60 cm thick external HDPE shield the neutron background can be reduced down to $0.24\pm 0.06$ events/year, lower than the background rate due to natural radioactivity (0.42 events/year), which was used as a benchmark for these calculations. The amount of shielding material needed can be significantly reduced by placing HDPE in the empty space between the lead shield and the steel vessel; in this way, it is sufficient to add 20 cm external HDPE shield to reduce the neutron background down to $0.15\pm0.05$ events/year.
      \rm}
\end{abstract}

\setcounter{footnote}{0}
\renewcommand{\thefootnote}{\arabic{footnote}}

\section{Introduction}

One of the most important problems in physics is related to the fundamental nature of neutrinos, namely whether they are Dirac or Majorana particles. 

Many experiments, currently under design or already running, will attempt to answer this question by looking for neutrinoless double beta decay: this lepton number-violating process is the "smoking gun" that would definitively prove the presence of a Majorana mass term for neutrinos. In a $\beta\beta$ decay, two neutrons are simultaneously transformed into protons, emitting two electrons and two antineutrinos. If they are Majorana particles, the two antineutrinos can annihilate each other, and all the energy of the decay would be carried out by the electrons, creating a bump at the end of the $\beta$ spectrum.

The No Neutrino Double-beta-decay Experiment (\NvDEx) is one of the experiments that will look for this kind of process~\cite{Nygren:2018ewr}. It will search for \0vbb~events using a high-pressure gas time projection chamber (TPC) filled with \SeF6, using \Se82 as a source of $\beta\beta$ decays. \NvDEx~will be placed at China Jinping Underground Laboratory (CJPL) which has the deepest rock shield in the world (2400 m rock overburden) and can provide an incredibly low-background environment. \NvDEx-100, which is currently under design \cite{NvDEx:2023zht}, will employ 100 kg of \SeF6; during the first phase natural Se will be used (\Se82 abundance: 8.7\%), later on, enriched Se (\Se82 abundance $>$90\%) will be employed. Due to the high electronegativity of \SeF6, free electrons in the gas will be captured almost immediately, which means that the negative particles drifting towards the high voltage plate will be ions and electron avalanche multiplication will not be possible. A new kind of sensor, Topemtal-S \cite{Gao:2019ohr,You:2021yqk}, has been developed and will be used in the detector, allowing it to reconstruct with high precision the energy of the events. The other main advantage of \NvDEx~is that the high Q-value of \Se82 (2.995 MeV) will place the Region of Interest (ROI) above most of the environmental background. 

The main challenge that neutrinoless double beta decay experiments must face is that the 0$\nu$2$\beta$ decay rate is proportional to $m_{\beta\beta}^2$, where $m_{\beta\beta}$ is the effective Majorana mass. However, since neutrino masses are incredibly small, this kind of process would be strongly suppressed. For this reason, an extremely low-background environment is a crucial requirement for the success of any neutrinoless double beta decay experiment~\cite{Henning:2016fad}: the detector should be carefully shielded in order to suppress the environmental background and great care should be taken to keep the contamination of the materials of the detector itself as low as possible. Moreover, a careful study of the background budget is required. 

There are several possible sources of background. $\alpha$ and $\beta$ particles created from radioactive decays can be stopped easily, so unless they are produced in the fiducial volume, they will not constitute a problem; $\gamma$'s, on the other hand, are more difficult to stop. They will be the main source of background for \NvDEx-100: it cannot be completely eliminated, because traces of radioactive elements will be present in the materials of the detector itself, however using a 20 cm lead shield it will be possible to suppress significantly the $\gamma$ flux coming from the experimental hall, reducing the background due to natural radioactivity down to 0.42 events in ROI/year\footnote{Please notice that in the computation of this source background, topological cuts have not been taken into account, they should be able to reduce even further the background rate (for example in the NEXT experiment they can reduce the $\gamma$ background by one order of magnitude \cite{NEXT:2012zwy}). For consistency, topological cuts will not be taken into account in this work as well, which means that the background rate here reported can be reduced even further.}~\cite{NvDEx:2023zht} or, equivalently, $1.4\times10^{-4}$ events/(kg$\cdot$keV$\cdot$y). Since adding additional shield will not reduce further this kind of background, such a value will be used in the current work as a benchmark.

Neutrons are more difficult to stop even than $\gamma$'s since they only interact weakly; they can be created from natural fission, ($\alpha$,n) processes and from cosmogenic muons. They can activate nuclei in the detector, even inside the fiducial volume itself, providing background. When they interact with a nucleus, they can produce $\gamma$'s via (n,$\gamma$) and (n,n'$\gamma$) reactions, or create unstable isotopes, for example via the reaction:
\begin{equation}
    ^{A}N+n\rightarrow \textrm{}^{A+1}N
\end{equation} 
$\gamma$'s could be dangerous if produced anywhere in the detector, while the electrons emitted via $\beta$ decays of activated nuclei could be a source of background only if the activation happened directly in the fiducial volume, since they are much easier to stop.

Cosmic rays are also related to several sources of background. When materials are manufactured on the ground, the cosmic rays will create unstable isotopes; most of them will decay almost immediately, however some of them will have relatively long half-lives, which means that, without any kind of cool-down period, they could still provide a non-negligible source of radioactivity even after the start of the experiment \cite{NvDEx:2023zht}.  Cosmogenic muons can travel deep underground. Interacting with the rocks surrounding the experimental hall, they can produce high-energy neutrons via spallation, moreover in principle they could even hit directly the fiducial volume. However at CJPL, due to the large rock overburden, the muon flux will be drastically suppressed, which means that this source of background can be neglected.

In this paper we will focus on the study of the neutron background. It is structured as follows: in Sec. \ref{sec:Simulations} we will present the simulations set-up, in Sec. \ref{sec:Isotopes} we will discuss which isotopes can be created in the fiducial volume and how they can contribute to the neutron background, while in Sec. \ref{sec:Gamma} we will examine the background due to neutron-induced $\gamma$'s. In Sec.~\ref{sec:Se82} we will discuss if and how using enriched Se can change the neutron-induced background and in Sec. \ref{sec:Results} we will summarize the results.

\section{Simulation Set-Up}
\label{sec:Simulations}

The neutron background was studied through Geant4 simulations \cite{Allison:2016lfl,Allison:2006ve,GEANT4:2002zbu}, using the Shielding Physics Lists, which allows the high precision treatment of low-energy neutrons; the G4NDL dataset was used for the low-energy neutron cross sections.

In an underground laboratory environment, neutrons can be produced from natural fission, ($\alpha$,n) reactions or from the interactions of cosmogenic muons with the rock surrounding the experimental hall. The latter could, in principle, be considerably harder to stop, since those spallation neutrons can reach significantly higher energies (up to a few GeV), however deep underground cosmogenic muons are strongly suppressed and the spallation neutron flux is several orders of magnitude smaller: the total neutron flux at CJPL is $(2.69\pm1.02)\times10^{-5}$cm$^{-2}$s$^{-1}$ \cite{Hu:2016vbu}, while the neutron flux from cosmogenic muons is estimated to be $8.37\times10^{-11}$cm$^{-2}$s$^{-1}$ \cite{121252}. In our simulations neutrons were created near the surface of the detector, assuming an isotropic distribution and with an energy spectrum equal to the one measured in CJPL and reported in \cite{Hu:2016vbu}, the total normalization factor was computed taking into account the total neutron flux. 

\begin{figure}[h!tbp]
\centering 
\includegraphics[width=0.6\textwidth,origin=c,angle=0]{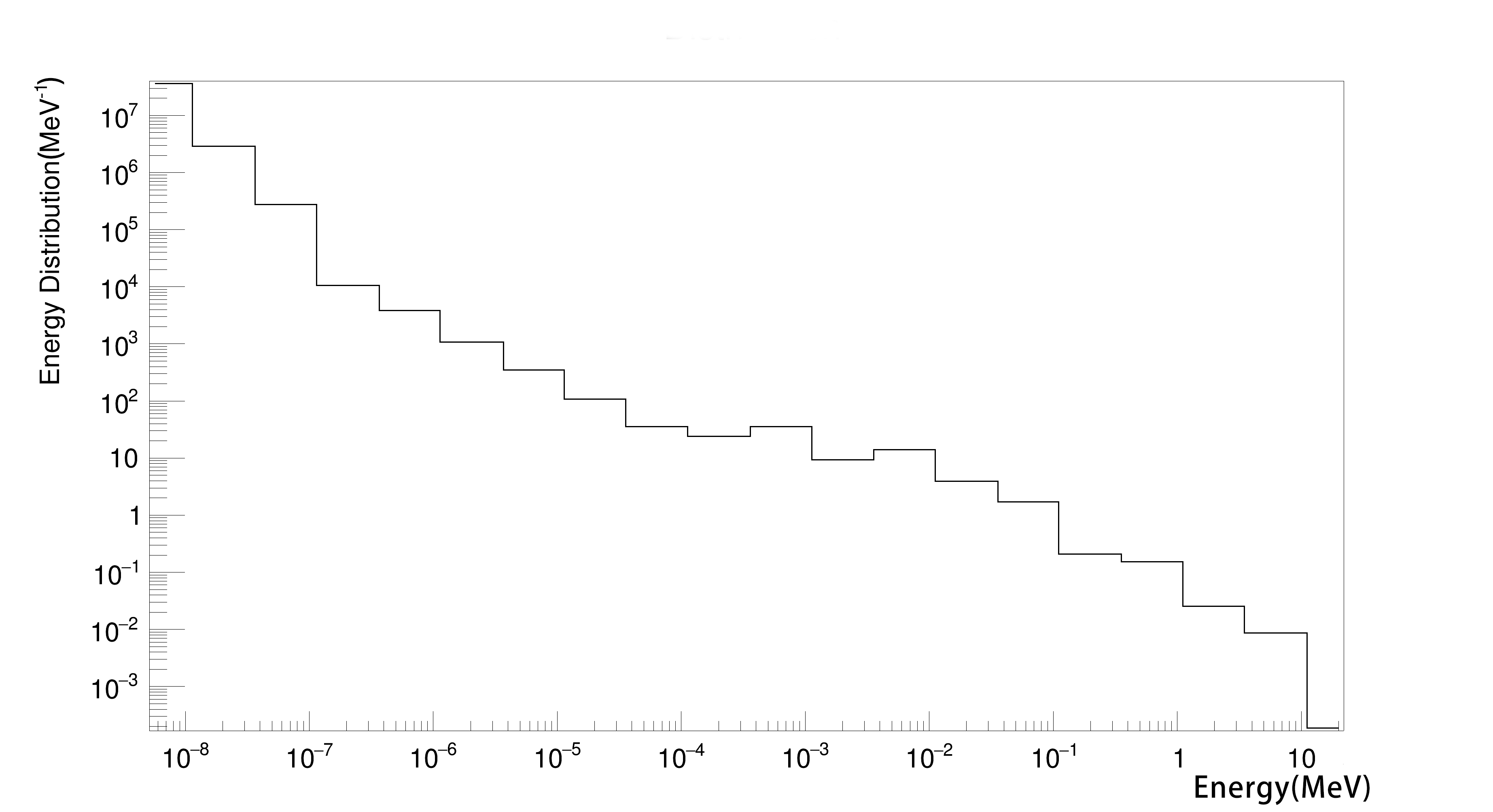}

\caption{\label{fig:ii} Neutron energy distribution used in simulations}
\end{figure}

\begin{figure}[htbp]
\centering 
\includegraphics[width=.8\textwidth]{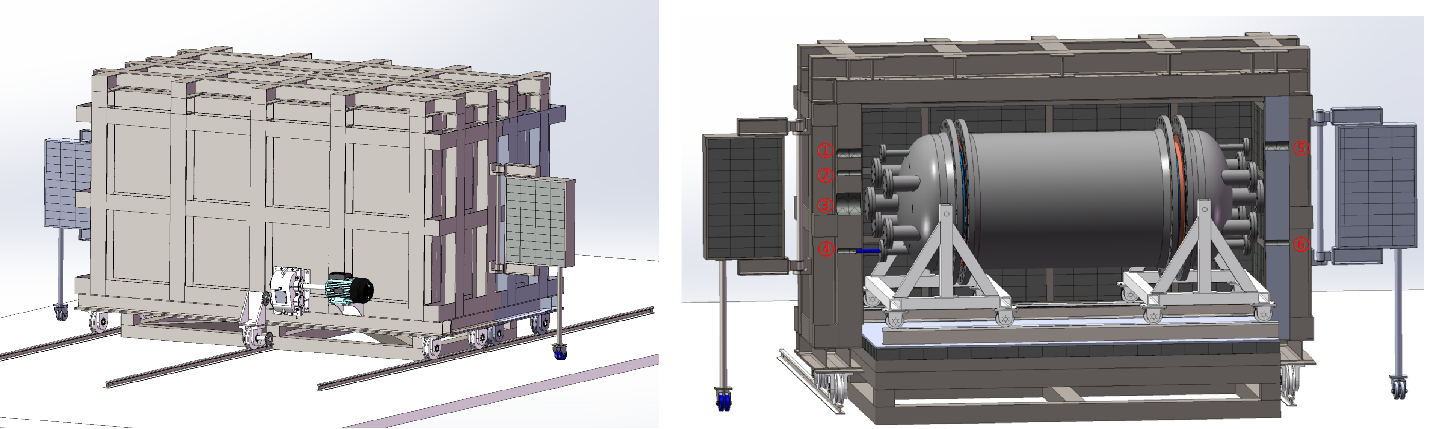}
\caption{\label{fig:Design} Preliminary design of \NvDEx~ detector }
\end{figure}
\begin{figure}[htbp!]
\centering 
\subfigure[No Shield]{
\begin{minipage}{0.4\linewidth}
\centering 
\includegraphics[width=0.9\textwidth,origin=c,angle=0]{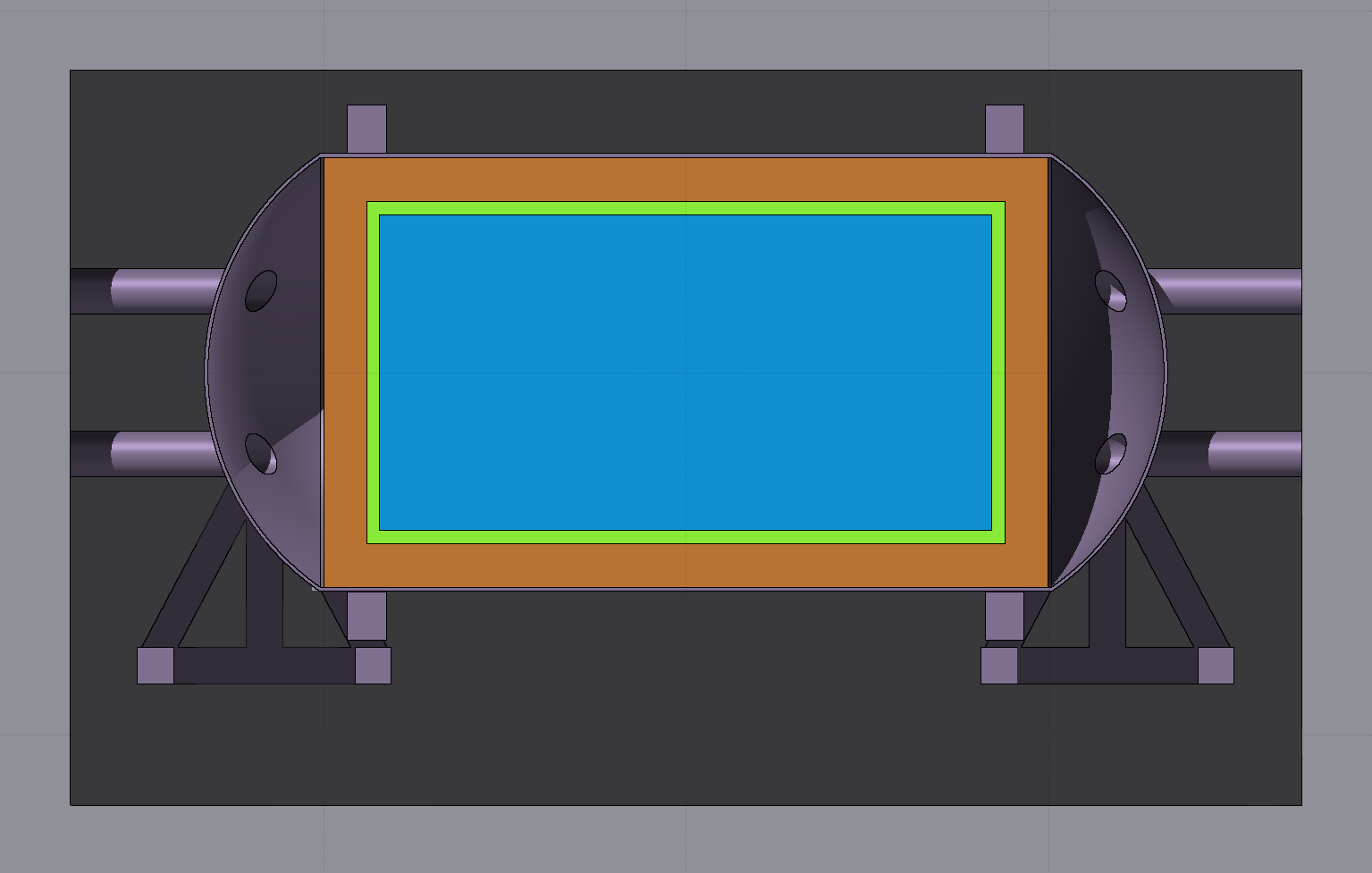}

\end{minipage}
}\\
\subfigure[External]{
\begin{minipage}{0.4\linewidth}
\centering 
\includegraphics[width=0.9\textwidth,origin=c,angle=0]{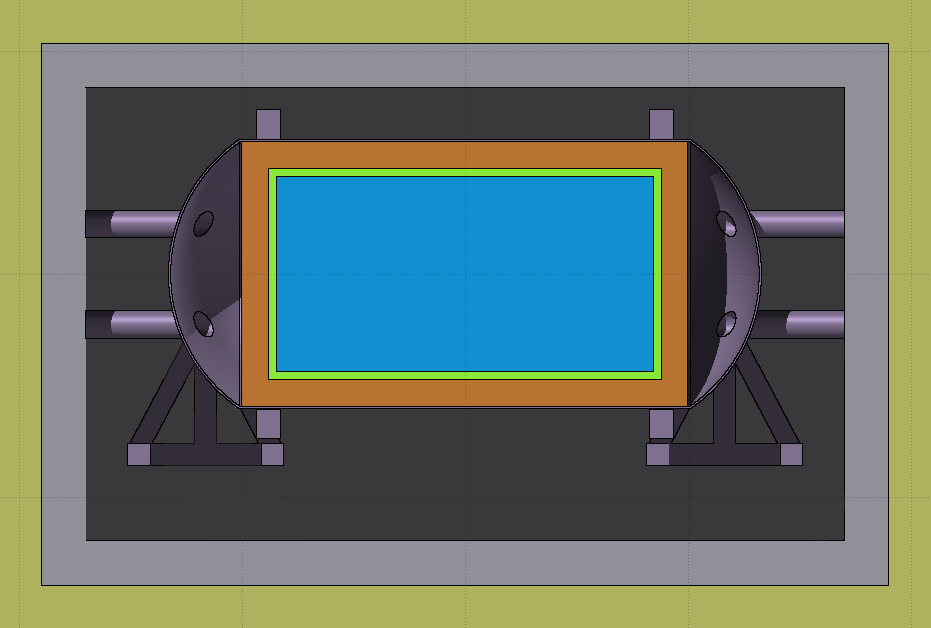}

\end{minipage}
}
\subfigure[Full]{
\begin{minipage}{0.4\linewidth}
\centering 
\includegraphics[width=0.9\textwidth,origin=c,angle=0]{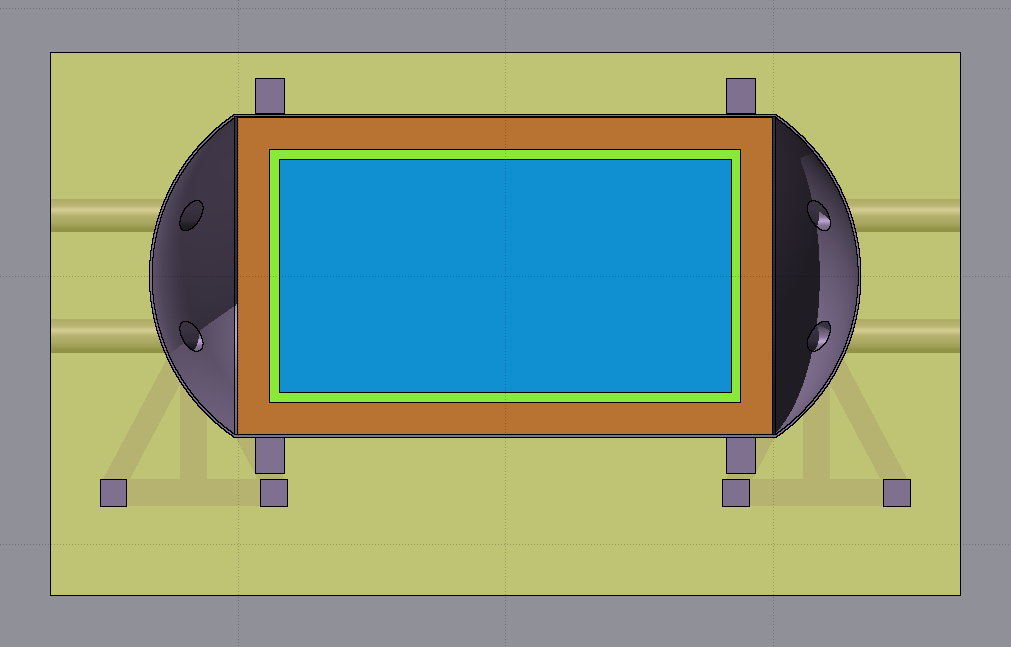}
\end{minipage}
}
\caption{Different configurations considered for the HDPE shield (top panel: no shield).}
\label{fig:Detector}
\end{figure}

A picture of the preliminary design of the \NvDEx~detector station can be seen in Fig.~\ref{fig:Design}, see Ref.~\cite{NvDEx:2023zht} for more details. In our simulations, we used the simplified design shown in the top panel of Fig.~\ref{fig:Detector}; it contains the lead castle shield (gray), Stainless Steel Vessel (SSV), the holders to support it and the pipes that will be connected to the outside of the shields (these three elements are in dark grey), as well as the Inner Copper Shield (ICS, orange) and the field cage (which was approximated as a POM shell inside the ICS, indicated in green in the picture) and the SeF$_6$ gas (blue). In the fisrt phase N$\nu$DEx will use natural Se (abundance of $^{82}$Se: 8.7\%), while in the second phase enriched Se will be used ($^{82}$Se abundance $>$90\%). We will use natural Se in our simulations; we will also show in Sec.~\ref{sec:Se82} that an increase in abundance of $^{82}$Se will only decrease the background rate.
The lead shield will stop the vast majority of the environmental $\gamma$'s, however in order to stop neutrons, low-Z materials are preferred. For this reason, high-density polyethylene (HDPE) shielding will be considered: it is formed by a chain of C$_2$H$_4$, the hydrogen atoms contained in it are very effective in slowing down and absorbing fast neutrons. Since it is relatively cheap and it is possible to obtain very low radioactive contamination levels, it is ideal for the task.

We considered two different configurations for the positioning of the HDPE shield, which are shown in the bottom panels of Fig.~\ref{fig:Detector}. The most straightforward solution is to place it outside the lead shield, this configuration will be indicated as "external". Another possibility is to fill the empty space between the SSV and the lead shield with HDPE, except for the regions occupied by the holder or the pipes (and, in case this is not sufficient, to add an external shield as well); this configuration will be indicated as "full" or "full filler". 

\section{ Neutron-Induced $\beta$ Decays}
\label{sec:Isotopes}


Neutrons can be captured by nuclei in the detector, creating unstable isotopes, that can emit $\alpha$ and $\beta$ particles during their decay.

If these particles are produced in the shields or in the supporting structures of the detector, it would not constitute a problem, since they will easily be stopped; however, if the nucleus activated is already inside the fiducial volume, its decay could be mistaken as a $\beta\beta$ event. 
In the \NvDEx~experiment, our working gas is SeF$_{6}$. The expected energy resolution is 1\% FWHM, and the ROI is between 2.98 MeV and 3.01 MeV~\cite{NvDEx:2023zht}; the only isotopes that could be a source of background are $^{20}$F, $^{16}$N, $^{19}$O and $^{83}$Se, which can be created via the reactions
\begin{eqnarray}
    ^{19}F+n\rightarrow ^{20}F \nonumber \\
    ^{19}F+n\rightarrow ^{19}O+p \nonumber \\
    ^{19}F+n\rightarrow ^{16}N+\alpha \nonumber \\
    ^{82}Se +n \rightarrow ^{83}Se
\end{eqnarray}
Other unstable isotopes, such as $^{79}$Se, $^{81}$Se, etc..., can be created as well, but they can be ignored since their Q-value is lower than the ROI.

In the decay of all these isotopes a single $\beta$ will be emitted, except for $^{83}$Se, which will have a decay chain:
\begin{equation}
    ^{83}Se\xrightarrow{}^{83}Br+e^{-}+\bar{\nu}  \qquad ^{83}Br\xrightarrow{}^{83}Kr+e^{-}+\bar{\nu}
\end{equation}
however the Q-value of $^{83}$Br decay is 0.977 MeV, lower than the ROI, moreover its half-life is 2.4 h, which means we do not have to worry about pile-up background\footnote{pile-up background could happen if two separate background events, both with energies $<$ROI, happen very close one to each other: they could be mistaken for a single event, and the sum of their energy could be within the ROI. See Ref.~\cite{NvDEx:2023zht} for an estimation of pile-up background in \NvDEx.}.

In Tab.~\ref{tab::Iso} are reported the Q-values and $P_{ROI}$, {\it i.e.} the fraction of $\beta$ emitted with energy within the ROI, for all the relevant isotopes.
 \begin{table}[htbp]
\centering
\caption{\label{tab::Iso} Q-value and $P_{ROI}$ for the relevant isotopes.}
\smallskip
\begin{tabular}{ccc}
\hline
Isotopes& Q-Value (MeV) & $P_{ROI}$\\
\hline
$^{20}F$ & 7.02 \cite{Tilley:1998wli} & $9.1\times10^{-3}$  \\ \hline
$^{16}N$ & 10.04 \cite{TILLEY19931} & $6.3\times10^{-3}$ \\  \hline
$^{19}O$ & 4.82 \cite{TILLEY19951} & $4.6\times10^{-3}$ \\  \hline
$^{83}Se$ & 3.67 \cite{McCutchan:2015vcl} & $2.4\times10^{-5}$ \\  \hline
\end{tabular}
\end{table}

\begin{figure}[h!]
\centering 
\includegraphics[width=0.6\textwidth,origin=c,angle=0]{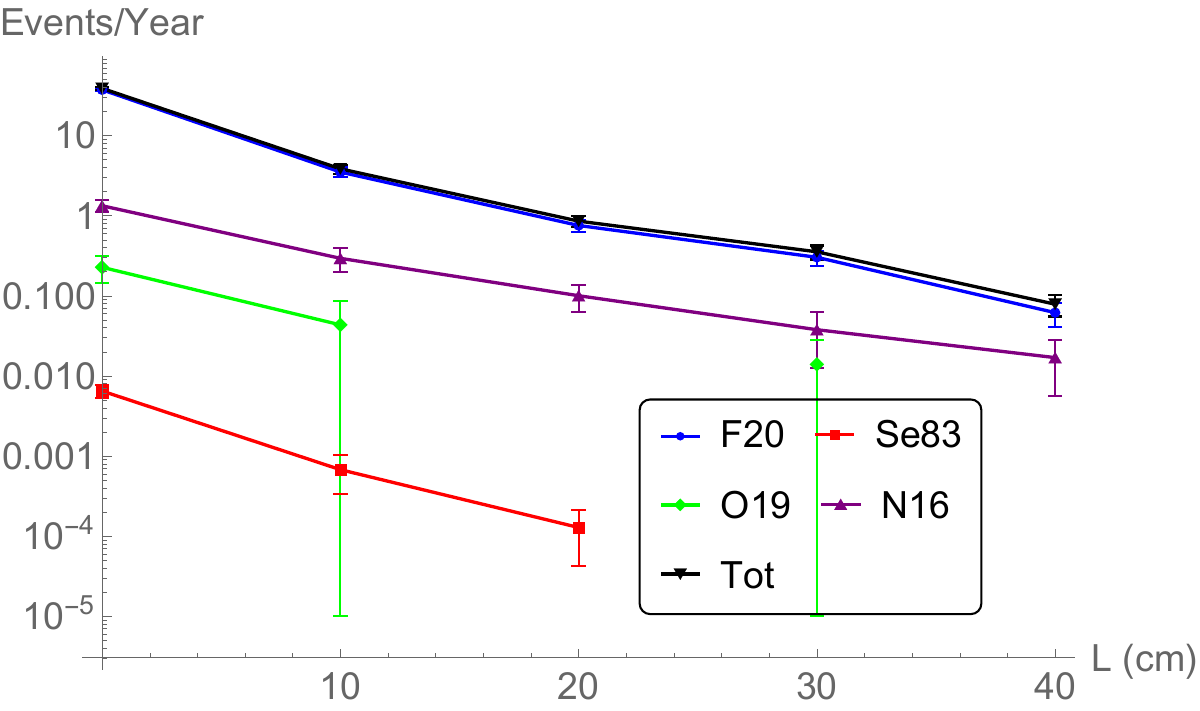}
\caption{\label{fig:IsotopesExternal} Contribution of each isotope to the background rate using external shield, as a function of the thickness}
\end{figure}

In Fig.~\ref{fig:IsotopesExternal} you can see the contribution of each isotope to the total background rate using external shield.
The main contribution to the background comes from $^{20}$F, which has the highest $P_{ROI}$, $9.1\times10^{-3}$. It should be pointed out that, after the decay, there could be  $\gamma$'s emitted as well (for example, in the decay of $^{20}$F, a 1.6 MeV $\gamma$ will be emitted), however those can be safely neglected in this particular case. First of all, the $\gamma$'s will travel considerably more in the detector before interacting (if they do interact at all), which means it will be easy to distinguish the tracks caused by the $\gamma$'s emitted during a radioactive decay from the ones due to the $\beta$ emitted in the same decay, so pile-up background is not a concern. 
Moreover, even if some high-energy $\gamma$'s can be emitted during these decays, these events are quite rare and their contribution to the total background rate will be negligible compared to the $\gamma$'s produced via (n,$\gamma$) and (n,n'$\gamma$) reactions (that can be created in the fiducial volume as well).

\begin{figure}[h!]
\centering 
\includegraphics[width=0.6\textwidth,origin=c,angle=0]{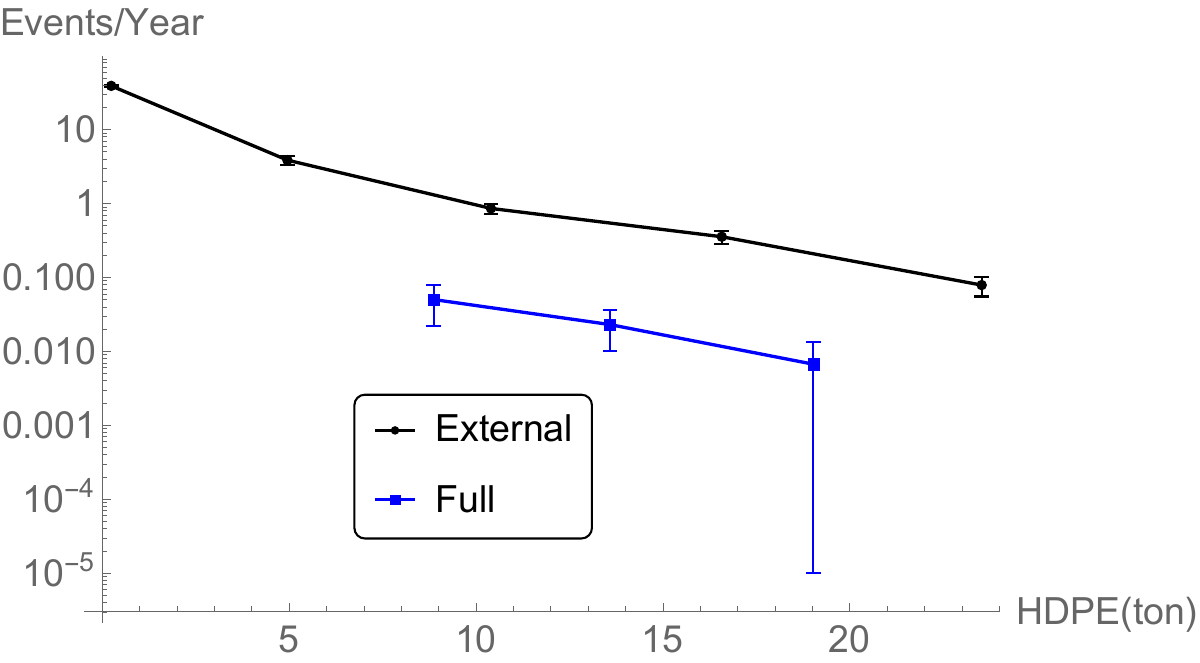}
\caption{\label{fig:IsotopesTotal} Background rate due to $\beta$ decays inside the fiducial volume the three configurations reported in Fig.~\ref{fig:Detector}. To facilitate the comparison, in the x-axes is reported the amount of HDPE used; the first data point corresponds to no external shielding at all, while in the others the thickness of the HDPE external shield is increased by 10 cm each step}
\end{figure}

We can also notice that the contribution of $^{83}$Se to the total background rate is negligible, since its $P_{ROI}$ is only $2.4\times10^{-5}$, almost three orders of magnitude lower than $^{20}$F's. $P_{ROI}$ for $^{16}$N and $^{19}$O are comparable to $^{20}$F's, but their production rates are considerably lower since these reactions have an energy threshold, namely the neutron energy must be higher than 0.5 and 3.5 MeV in order to create $^{16}$N and $^{19}$O, respectively. In Fig.~\ref{fig:IsotopesTotal} it is reported the total background rate due to $\beta$ decays inside the fiducial volume for all the configurations considered; to facilitate the comparison, in the x-axis it is reported the total amount of HDPE used, not the thickness of the external shield.

\section{Neutron-Induced $\gamma$'s}
\label{sec:Gamma}

Environmental neutrons can create $\gamma$'s while interacting with the material of the detector, via (n,$\gamma$) or (n, n'$\gamma$) reactions. For example, if a hydrogen atom absorbs a thermal neutron, it will emit a characteristic 2.2 MeV $\gamma$. Since this energy is lower than the ROI, these $\gamma$'s cannot provide background, however all the other elements present in the materials of the detector will emit higher-energy $\gamma$'s: their energies can go up to 5 MeV if a neutron is captured by a C atom (present in HDPE), 7-9 MeV if the target is Cu, Pb or Fe. 

\begin{figure}[htbp!] 
{
\centering
\includegraphics[width=0.9\textwidth,origin=c,angle=0]{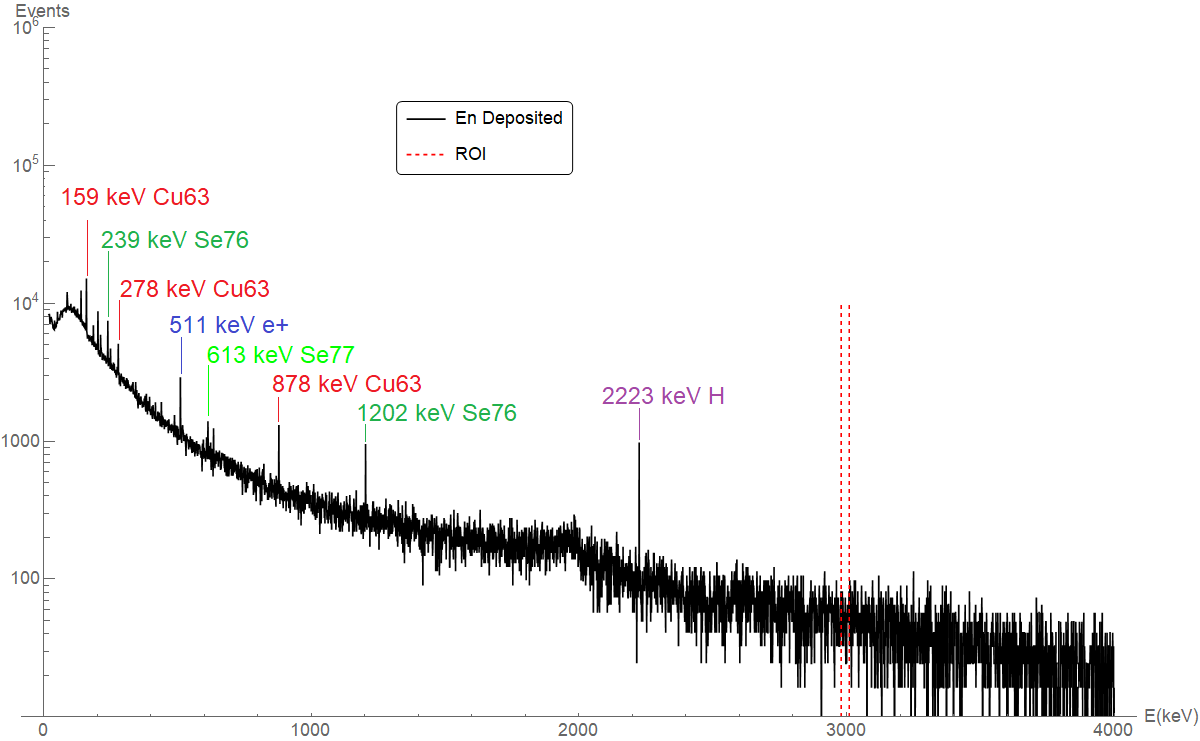}
}

\caption{Energy deposited in the detector by neutron-induced $\gamma$'s, if no HDPE shielding is added. Between red dashed lines it is indicated the ROI} 
\label{fig:EnergyDep}
\end{figure}

In Fig.~\ref{fig:EnergyDep} it is reported the energy deposited in the detector by neutron-induced $\gamma$'s, in the absence of HDPE shielding. It is possible to see several peaks due to neutron capture: the 2.2 MeV peak due to hydrogen capture is clearly visible, as well as the peaks at 159, 279 and 878 keV due to $^{63}$Cu capture. It is also possible to see peaks at 239 and 1202 keV, due to neutron capture by $^{76}$Se and at 613 keV due to $^{77}$Se \cite{10.1007/978-3-642-58113-7_227}. The peak at 511 keV is not due to neutron capture, but to electron-positron annihilation. 

\begin{figure}[htbp!] 
{
\centering
\includegraphics[width=0.9\textwidth,origin=c,angle=0]{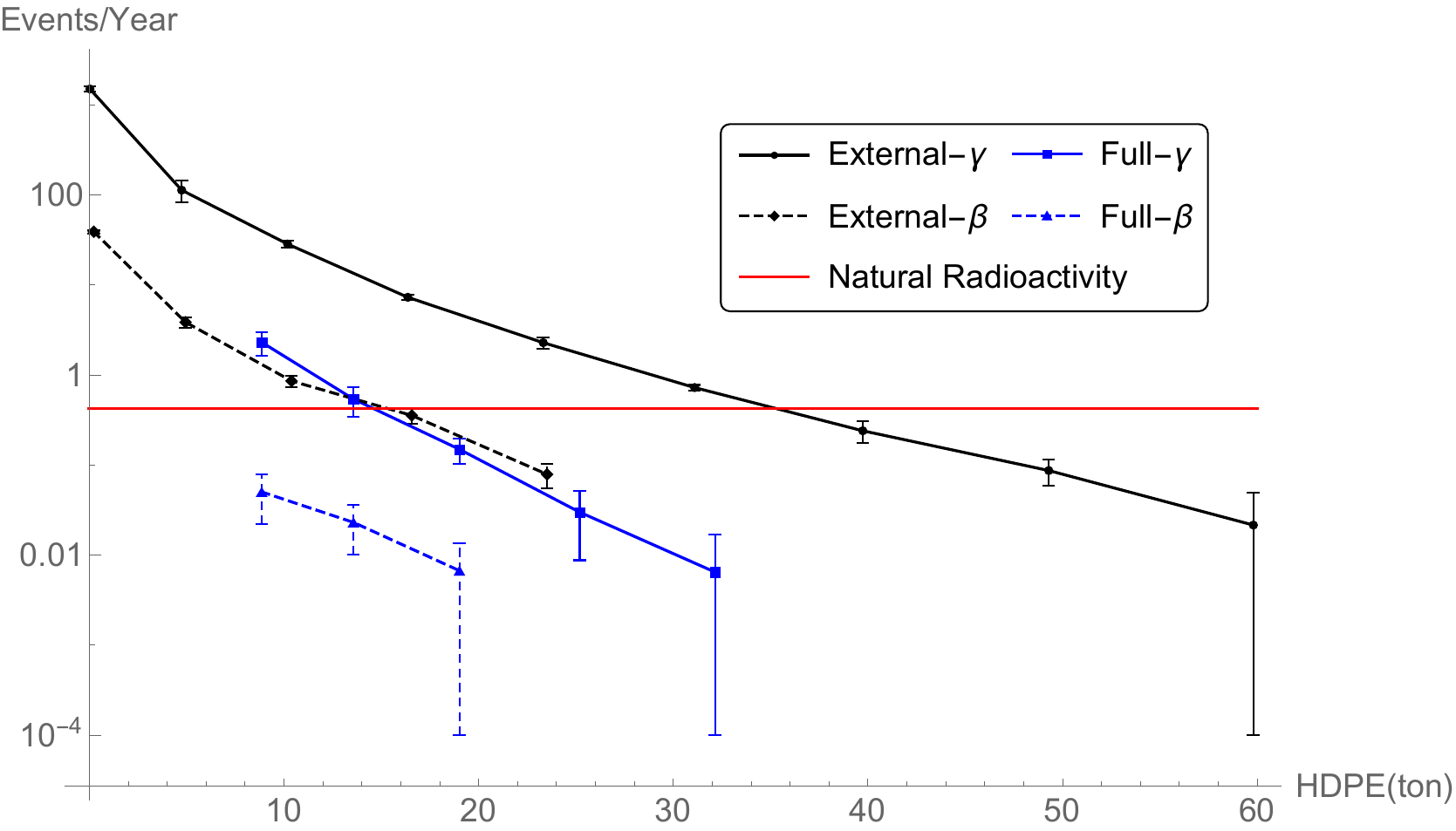}
}
\caption{\label{fig::TotBackgroundGamma} Background due to neutron-induced $\gamma$'s and $\beta$ decays. For comparison, the background due to natural radioactivity is indicated as well, as estimated in \cite{NvDEx:2023zht} }
\label{fig:3c}
\end{figure}

In Fig.~\ref{fig::TotBackgroundGamma} it is reported the background rate due to neutron-induced $\gamma$'s; for comparison, the background rates due to neutron-induced $\beta$ decay and to natural radioactivity are reported as well. It is possible to see that, in all configurations, the $\gamma$ contribution is dominant. If no HDPE shield is used, there would be 1492$\pm$93 events in ROI/year due to neutron-induced $\gamma$'s, while the background due to $\beta$ decay in the fiducial volume would be only 38.7 $\pm$ 1.6 events/year. Moreover, even if HDPE is placed between the lead shield and the SSV, the neutron-induced background would still be larger than the one due to natural radioactivity, which means that an external shield will always be needed. The presence of a filler between the lead shield and the SSV, however, will reduce significantly the amount of HDPE needed. If only an external shield is present, it must be at least 60 cm thick, which would require around 40 tons of HDPE. On the other hand, using a filler, a 20 cm external shield will be sufficient to reduce the neutron background to subdominant levels; this would correspond to 19 tons of HDPE in total.  

  \section{Enriched Se}
 \label{sec:Se82}
 
After the first phase, where natural Se will be employed, N$\nu$DEx will use enriched Se, where the abundance of $^{82}$Se will be increased up to 90\%. As it can be seen in Tab.~\ref{tab:SeCrossSetion}, $^{82}$Se has the lowest absorption cross-section for thermal neutrons. In particular, $^{76}$Se and $^{77}$Se, have absorption cross-sections which are three orders of magnitude higher.
This means that, if the abundance of $^{82}$Se increases, fewer neutrons will be absorbed by Se isotopes, which will have two consequences: on one hand, the production rate of other isotopes, in particular $^{20}$F, would increase; on the other hand, there would be fewer $\gamma$'s emitted due to neutron capture in the fiducial volume.
\begin{table}[h]
\centering
\caption{\label{tab:SeCrossSetion} Thermal neutron absorption cross section for Se isotopes (cross sections taken from \cite{doi:10.1080/10448639208218770})}
\smallskip
\begin{tabular}{ccc}
\hline
Isotope& Abundance (\%)& $\sigma_n$ (barn)\\
\hline
$^{74}$Se & 0.89 & 51.8 \\ \hline
$^{76}$Se & 9.4 & 85. \\ \hline
$^{77}$Se & 7.6 & 42. \\ \hline

$^{78}$Se & 23.7 & 0.43 \\ \hline
$^{80}$Se & 49.6 & 0.61 \\ \hline
$^{82}$Se & 8.7 & 0.044 \\ \hline
\hline
\end{tabular}
\end{table}

\begin{figure}[htbp!]
\centering 
\begin{minipage}{0.45\linewidth}
\centering 
\includegraphics[width=1\textwidth,origin=c,angle=0]{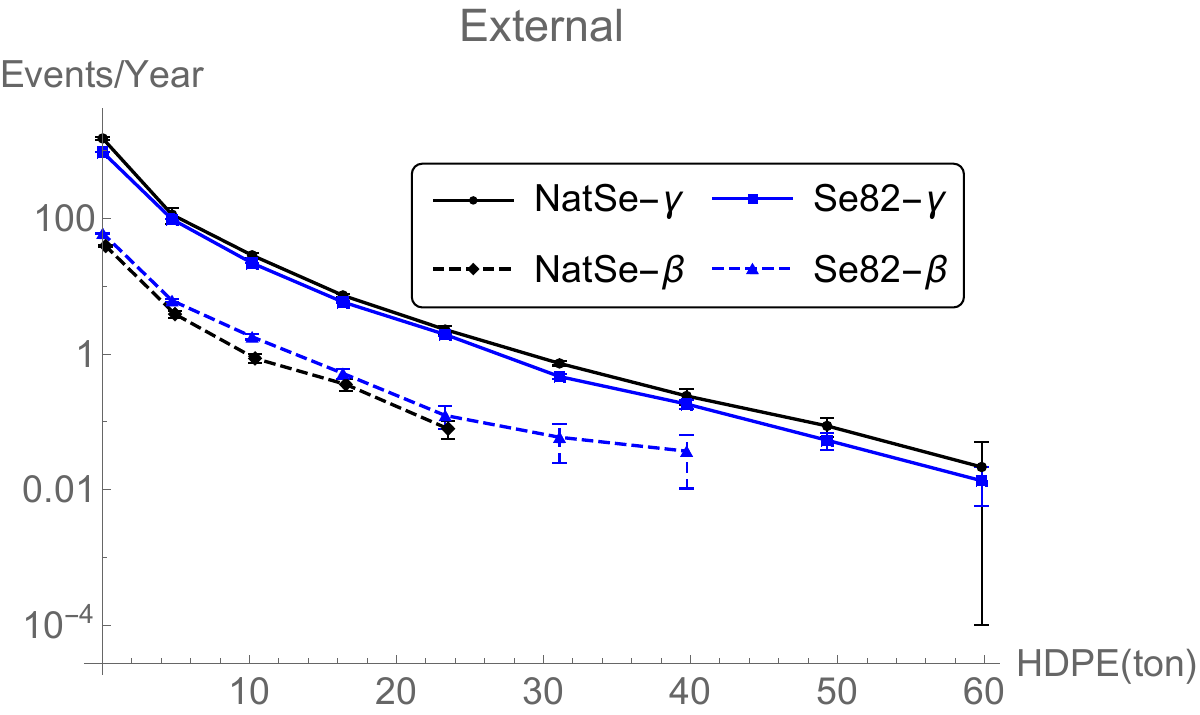}

\end{minipage}
\begin{minipage}{0.45\linewidth}
\centering 
\includegraphics[width=1\textwidth,origin=c,angle=0]{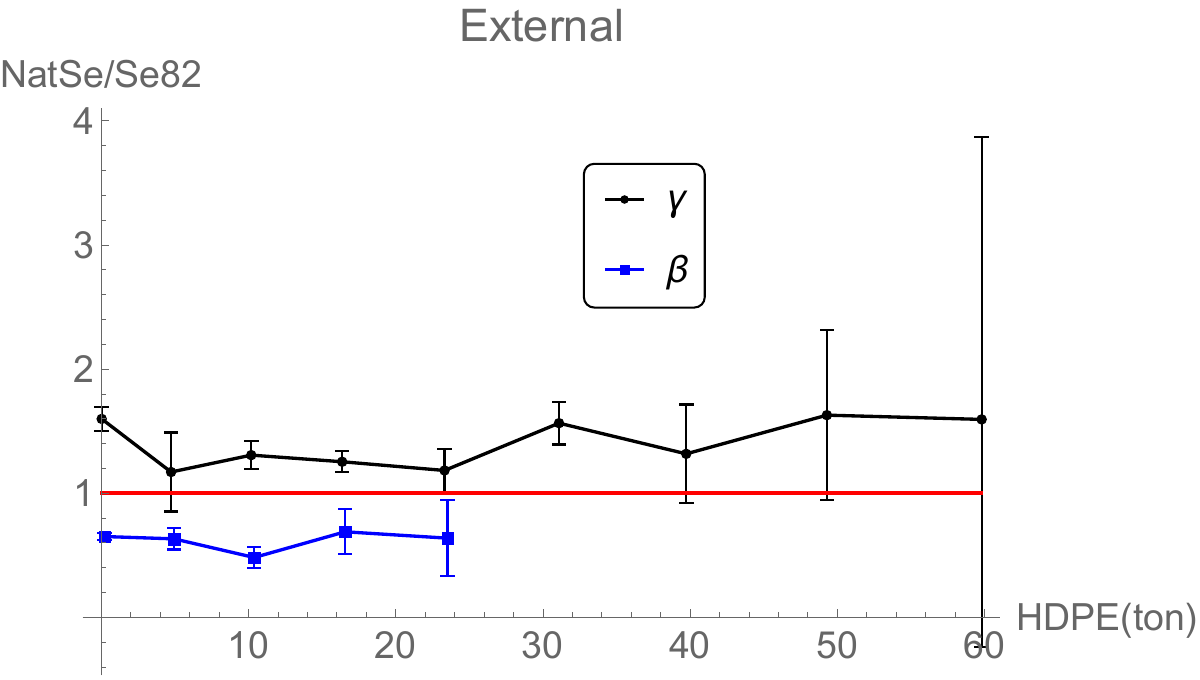}
\end{minipage}
\caption{\label{fig:ComparisonSe82}
 Left panel: background rate (external shield) using natural Se or $^{82}$Se. Right panel: ratio between the background rates }
\end{figure}

In Fig.~\ref{fig:ComparisonSe82} (left panel) it is reported the background rate using natural Se or pure $^{82}$Se \footnote{Since the exact isotopic composition of enriched Se is not known yet, we have used pure $^{82}$Se in our simulations, as the most extreme case: the actual background rate for enriched Se would lie in between those results and the ones obtained with natural Se} (only external shield was considered). We can see that, as expected, the neutron-induced $\gamma$ background is lower if the gas contains only $^{82}$Se. Such a difference is of the order of 50\% if there is no shielding at all (Fig.~\ref{fig:ComparisonSe82}, right panel), it is more difficult to give a precise estimation if the neutron background is strongly suppressed, due to the large statistical errors\footnote{it should be pointed out that a more precise estimation of the background rate would be not useful for the purpose of this paper since we are interested only in an upper limit on the neutron background}. The background due to neutron-induced $\beta$ decays is higher if $^{82}$Se is used, but since such a contribution is strongly subdominant, the total background is nonetheless lower.  

 \section{Conclusions}
 \label{sec:Results}

In this paper we used Geant4 simulations to study the neutron background for the \NvDEx~experiment.

The main source of background are neutron-induced $\gamma$'s. Neutrons could also activate nuclei directly in the fiducial volume and induce $\beta$ decays, however this source of background is strongly subdominant.

The material we considered for shielding is HDPE: due to its high hydrogen content, it is very effective in slowing down and stopping neutrons. If only an external HDPE shield is used, it must be 60 cm thick for the neutron background to be reduced down to $0.24\pm 0.06$, lower than the natural radioactivity background (0.42 events/year); the total amount of HDPE needed in this configuration would be of the order of 40 ton.

We found that the amount of HDPE needed can be significantly decreased if an HDPE is placed between the lead shield and the SSV.
A certain amount of external shield is always needed, however the requirements on the thickness of such a shield can be significantly lowered. Using an HDPE filler, only a 20 cm thick external shield will be required to reduce the neutron-induced background down to $0.15\pm0.05$ events/year, which corresponds to 19 tons of HDPE. 

We have also shown that, if enriched Se is used, the neutron background rate will be lower. Indeed, if the abundance of $^{82}$Se increases, the total neutron absorption cross-section of Se decreases. On one hand, this means that the production rate of other isotopes, in particular $^{20}$F, will increase, however this source of background is strongly subdominant. On the other hand, fewer neutron-induced $\gamma$'s will be created in the fiducial volume: since this is the main source of neutron background, the total  background rate will decrease.

\section* {Acknowledgement}
This project is supported by the National Key Research and Development Program of China 2021YFA1601300, From-0-to-1 Original Innovation Program of Chinese Academy of Sciences ZDBS-LY-SLH014, and International Partner Program of Chinese Academy of Sciences GJHZ2067. We wish to thank Hao Qiu, Qiang Hu and Fengyi Zhao for the useful discussions and suggestions.








\printbibliography

@article{Henning:2016fad,
    author = "Henning, Reyco",
    title = "{Current status of neutrinoless double-beta decay searches}",
    doi = "10.1016/j.revip.2016.03.001",
    journal = "Rev. Phys.",
    volume = "1",
    pages = "29--35",
    year = "2016"
}

@article{Nygren:2018ewr,
    author = "Nygren, D. R. and Jones, B. J. P. and L\'opez-March, N. and Mei, Y. and Psihas, F. and Renner, J.",
    title = "{Neutrinoless Double Beta Decay with $^{82}$SeF$_6$ and Direct Ion Imaging}",
    eprint = "1801.04513",
    archivePrefix = "arXiv",
    primaryClass = "physics.ins-det",
    doi = "10.1088/1748-0221/13/03/P03015",
    journal = "JINST",
    volume = "13",
    number = "03",
    pages = "P03015",
    year = "2018"
}

@article{Hu:2016vbu,
    author = "Hu, Qingdong and others",
    title = "{Neutron background measurements at China Jinping underground laboratory with a Bonner Multi-sphere Spectrometer}",
    eprint = "1612.04054",
    archivePrefix = "arXiv",
    primaryClass = "physics.ins-det",
    doi = "10.1016/j.nima.2017.03.048",
    journal = "Nucl. Instrum. Meth. A",
    volume = "859",
    pages = "37--40",
    year = "2017"
}

@article{121252,
title = "Monte Carlo simulation of muon radiation environment in China Jinping Underground Laboratory",
journal = "High Power Laser and Particle Beams",
volume = "24",
number = "121252",
pages = "3015",
year = "2012",
note = "",
issn = "1001-4322",
doi = "10.3788/HPLPB20122412.3015",
url = "http://www.hplpb.com.cn/en/article/doi/10.3788/HPLPB20122412.3015",
author = "Su Jian and Zeng Zhi and Liu Yue and Yue Qian and Ma Hao and Cheng Jianping"
}

@misc{NvDEx:2023zht,
    author = "Cao, X. and others",
    collaboration = "NvDEx",
    title = "NvDEx-100 Conceptual Design Report",
    archivePrefix = "arXiv",
    eprint = "2304.08362",
    primaryClass = "physics.ins-det",
    month = "4",
    year = "2023"
}

@article{Allison:2016lfl,
    author = "Allison, J. and others",
    title = "{Recent developments in Geant4}",
    reportNumber = "FERMILAB-PUB-16-447-CD",
    doi = "10.1016/j.nima.2016.06.125",
    journal = "Nucl. Instrum. Meth. A",
    volume = "835",
    pages = "186--225",
    year = "2016"
}

@article{Allison:2006ve,
    author = "Allison, John and others",
    title = "{Geant4 developments and applications}",
    reportNumber = "SLAC-PUB-11870",
    doi = "10.1109/TNS.2006.869826",
    journal = "IEEE Trans. Nucl. Sci.",
    volume = "53",
    pages = "270",
    year = "2006"
}

@article{GEANT4:2002zbu,
    author = "Agostinelli, S. and others",
    collaboration = "GEANT4",
    title = "{GEANT4--a simulation toolkit}",
    reportNumber = "SLAC-PUB-9350, FERMILAB-PUB-03-339, CERN-IT-2002-003",
    doi = "10.1016/S0168-9002(03)01368-8",
    journal = "Nucl. Instrum. Meth. A",
    volume = "506",
    pages = "250--303",
    year = "2003"
}

@article{NEXT:2012zwy,
    author = "Alvarez, V. and others",
    collaboration = "NEXT",
    title = "{NEXT-100 Technical Design Report (TDR): Executive Summary}",
    eprint = "1202.0721",
    archivePrefix = "arXiv",
    primaryClass = "physics.ins-det",
    doi = "10.1088/1748-0221/7/06/T06001",
    journal = "JINST",
    volume = "7",
    pages = "T06001",
    year = "2012"
}

@article{Gao:2019ohr,
    author = "Gao, Chaosong and An, Mangmang and Huang, Guangming and Huang, Xing and Mei, Yuan and Sun, Quan and Sun, Xiangming and Xiao, Le and Yang, Ping",
    title = "{A Low-Noise Charge-Sensitive Amplifier for Gainless Charge Readout in High-Pressure Gas TPC}",
    doi = "10.22323/1.343.0083",
    journal = "PoS",
    volume = "TWEPP2018",
    pages = "083",
    year = "2019"
}

@article{You:2021yqk,
    author = "You, Bihui and Xiao, Le and Sun, Xiangming and Mei, Yuan and Huang, Guangming",
    title = "{A distributed readout network ASIC for high-density electrode array targeting at neutrinoless double-beta decay search in a Time Projection Chamber}",
    doi = "10.1016/j.nima.2020.164871",
    journal = "Nucl. Instrum. Meth. A",
    volume = "988",
    pages = "164871",
    year = "2021"
}

@article{Tilley:1998wli,
    author = "Tilley, D. R. and Cheves, C. M. and Kelley, J. H. and Raman, S. and Weller, H. R.",
    title = "{Energy levels of light nuclei, A = 20}",
    doi = "10.1016/S0375-9474(98)00129-8",
    journal = "Nucl. Phys. A",
    volume = "636",
    pages = "249-364",
    year = "1998"
}

@article{TILLEY19931,
author = "D.R. Tilley and H.R. Weller and C.M. Cheves",
title = "Energy levels of light nuclei A = 16–17",
doi = "10.1016/0375-9474(93)90073-7",
journal = "Nuclear Physics A",
volume = "564",
number = "1",
pages = "1-183",
year = "1993",
}

@article{McCutchan:2015vcl,
    author = "McCutchan, E. A.",
    title = "{Nuclear Data Sheets for A = 83}",
    doi = "10.1016/j.nds.2015.02.002",
    journal = "Nucl. Data Sheets",
    volume = "125",
    pages = "201--394",
    year = "2015"
}

@article{TILLEY19951,
author = "D.R. Tilley and H.R. Weller and C.M. Cheves and R.M. Chasteler",
title = "Energy levels of light nuclei A = 18–19",
doi = "https://doi.org/10.1016/0375-9474(95)00338-1",
journal = "Nuclear Physics A",
volume = "595",
number = "1",
pages = "1-170",
year = "1995",
}

@article{doi:10.1080/10448639208218770,
author = " Varley F.   Sears ",
title = "Neutron scattering lengths and cross sections",
journal = "Neutron News",
volume = "3",
number = "3",
year  = "1992",
note=" Data retrieved from the NIST Center for Neutron Research, available at https://www.ncnr.nist.gov/resources/n-lengths/ (accessed on July 17, 2023)"
}

@article{10.1007/978-3-642-58113-7_227,
author="Bhat, M. R.",
editor="Qaim, Syed M.",
title="Evaluated Nuclear Structure Data File (ENSDF)",
journal="Nuclear Data for Science and Technology",
year="1992",
note=" Data retrieved from the National Nuclear Data Center (NNDC) using CapGam, available at https://www.nndc.bnl.gov/capgam/ (accessed on July 17, 2023)"
}
\end{document}